\documentclass[11pt]{article}
\usepackage{cospar}
\usepackage{url}
\usepackage{times}

\usepackage{graphicx}
\usepackage[figuresright]{rotating}

\usepackage[sectionbib]{natbib} 
\title{HIGH-ENERGY RADIATION FROM PULSARS : A THREE DIMENSIONAL MODEL APPROACH}
\author{K.S. Cheng\address{Department of Physics, University of Hong Kong, Hong Kong, China}}
\begin{document}
\maketitle

\begin{abstract}
We suggest that some observational features of high-energy radiation from pulsars should be explained in terms of three dimensional geometric models, e.g.~the phase-resolved X-ray and $\gamma$-ray spectra and the energy dependent light curves from various pulsars. In this paper, we present a three dimensional pulsar outer-magnetospheric gap model to explain these observational features. The outer-magnetospheric gap is proposed to form near the null charged surface and extend toward the light cylinder. The other two geometric dimensions of the outer-magnetospheric gap, i.e.~the vertical size and the azimuthal extension can be determined self-consistently. 
We apply this model to explain the observed phase-dependent spectra and the energy-dependent light curves of various pulsars.
\end{abstract}

\section{INTRODUCTION}

The non-thermal high-energy radiation from rotation-powered pulsars is believed to be emitted from their magnetosphere. In the past one decade, there has been a lot of progress in detecting and understanding high-energy radiation from rotation powered pulsars. In particular, the X-ray and $\gamma$-ray data obtained by the satellite observatories ROSAT, ASCA, RXTE, BeppoSAX, RXTE, CGRO, Chandra and XMM-Newton provide very important constraints for the theoretical models, which are used to explain the high-energy radiation from rotation powered pulsars. The observed data is so rich that the local properties in the magnetosphere, e.g.~retarded relativistic effect, the time of flight, the strength of the local magnetic field, the local radius of curvature, the local soft photon density etc., are necessarily to be used in order to explain the observed phase-dependent spectra and energy-dependent light curves from pulsars. These local properties associated with the observed photons not only depend on the global pulsar parameters, i.e.~rotation period and surface magnetic field, but also depend on the angle between the magnetic axis and the rotation axis (the inclination angle $\alpha$) as well as the viewing angle ($\zeta$).

EGRET has observed the phase-resolved emission characteristics such as pulse profiles and phase-resolved spectra of Crab, Vela and Geminga Pulsars (Thompson et al.~1996). Fierro et al.~(1998) have divided the observed photons of the Crab pulsar in the energy range from 100 MeV to 10 GeV into eight different phase bins and have shown that the spectra of each phase bin is very different. If a power-law fitting is used, the spectral indices of these phase bin can vary from -1.71 to -2.65. The observations of X-ray emission from the Crab pulsar also indicate that its spectral energy distribution is changing within its double peak pulse
profile. Pravdo et al.~(1997) analyzed the RXTE data between
5 and 250 keV, and showed an evolution of the spectral index across the
X-ray pulse in a reverse S shape. Massaro et al.~(2000) presented the phase-resolved analysis results based on BeppoSAX data. Kuiper et al.~(2001) have summarized the basic observational properties of the Crab pulsar from soft X-rays to high-energy gamma-rays. Here, we summarize the basic observed
properties at X-rays are as follows: (i) a double pulse profile with a bridge
separated by $\sim 0.4$ in phase has been observed; (ii) the spectra of the
two peaks soften and the softest spectrum is the first peak; (iii) the
spectrum the peaks hardens; and (iv) spectral
indices are clearly increasing with energy over all the phase interval.
The Chandra X-Ray Observatory has also observed the Crab pulsar using the Low-Energy Transmission Grating with the High-Resolution Camera. Time-resolved zeroth-order images reveal the pulsar emits X-rays at all pulse phases (Tennant et al.~2001). A preliminary analysis of the dispersed data indicates that the spectral indices evolve as a function of pulse phase (Weisskopf, 2002). The Chandra result is in semi-quantitative agreement with previous measurements (e.g.~Pravdo et al.~1997; Massaro et al.~2000) at various energies. However, the Chandra results extend the phase coverage through pulse minimum. In summary, observations strongly suggest that these phase-dependent properties come from different parts of the pulsar magnetosphere.
The question is how to explain theoretically the phase dependent 
properties of the observed high-energy radiation from the Crab pulsar as well as from other gamma-ray pulsars.
We suggest that the above examples indicate the necessity of using three dimensional outer magnetospheric gap (hereafter outer gap) models to explain the observed data.

\section{THREE DIMENSIONAL OUTER GAP MODELS}

The outer gaps, powerful acceleration regions, can form in the vicinity of
``null charge surface" (${\bf{\Omega}}\cdot{\bf{B}}=0$) (Holloway, 1973;
Cheng et al.~1976) because the charged carriers on each
side of the null charge surface have opposite charges. In fact, the
charge density of the magnetosphere in the  corotating frame of a neutron
star is (Goldreich and Julian, 1969) $\rho_0\sim
-({\bf{\Omega}}\cdot{\bf{B}}/ 2\pi c)$, where ${\bf{B}}$ and
${\bf{\Omega}}$ are the magnetic field and angular velocity of the
neutron star. The charge density will change sign when a global current
flows through the null surface where ${\bf{\Omega}}\cdot{\bf{B}}=0$. As a
result, a charge-deficient region ($\rho\approx 0$) in the outer
magnetosphere near the null surface will be formed. Any deviation of
the charge density from $\rho_0$ results in an electric field along {\bf{B}}. 
Cheng et al.~(1986a,b, hereafter CHRI and CHRII respectively) argue that this electric field can become strong enough to accelerate $e^{\pm}$
pairs to ultra relativistic energies. These $e^{\pm}$ pairs could radiate
$\gamma$-ray tangential to the curved {\bf{B}} field lines. These
``curvature $\gamma$-rays" are further converted into $e^{\pm}$ pairs via
$\gamma + \gamma\rightarrow e^+ + e^-$. Therefore, in order to keep a
steady state current flow and the charge density $\rho_0$ in the regions
outside the gap, the gap will grow until it is large enough and the
electric field is strong enough to maintain a copious supply of charges
to the rest of the open field line region. If the gap ends in a region
$\rho_0\neq 0$, charges from the surrounding region will flow in through
the end. If both ends are located on the null surface, any $e^{\pm}$
produced in the gap will act to replace the charge deficiency inside the
gap, and finally the gap will be filled up. However, if a vacuum gap
extends to the light cylinder, charged particles created in the gaps will
escape from the magnetosphere, so the gap will not be quenched. Hence,
stable outer gaps (if they exist) are those from the null surface 
to the light cylinder along the last closed field lines. In each outer
gap, the inner boundary of the outer gap lies near the intersection of the
null surface where $\rho=0$ and the boundary of the closed  field lines
of the star on which the magnetosphere current does not flow. The
thickness of the outer gap is bounded from above by a layer of electric
current which contributes a surface charge density. 

According to CHR model, four outer gaps exist in the open zone in the plane of
(${\bf{\Omega}}$, ${\bf{\mu}}$) (two of them are topologically connected
in three dimensional space), but only two longer outer gaps should give
observed fan beams. They argued that these two longer outer gaps may
create enough $\gamma$-ray and $e^{\pm}$ pairs to quench the two shorter,
less powerful ones. In the CHR model, the $\gamma$-ray emission is
approximated to occur only along the last closed field line in the plane
of the dipole and rotation axes. Because the charged particles of both
positive and negative charges are accelerated in the gap, the emissions
should beam both toward and away from the pulsar. Therefore, the observed
fan beams consist of those coming from different gaps and the measured phase
separation between the two peaks is determined by the time travel
difference between these two outer gaps, relativistic aberration of
emission and the bending of the magnetic field lines near the light
cylinder. The emission from each peak is highly cusped because of the
relativistic aberration, so there will be some bridge emission but very
little other offpulse emission. Obviously, the pulse profiles of CHR model
are not consistent with the observed those. So the three dimensional 
description of the outer gaps is necessary.

After studying the $\gamma$-ray production and light curves for various
magnetosphere geometries based on the CHR model, Chiang and Romani (1992)
assumed that gap-type regions could be supported along all field lines
which define the boundary between the closed region and open field line
region rather than just on the bundle of field lines lying in the plane containing the rotation and magnetic dipole axes. In this case, photons
are generated which travel tangential to the local magnetic 
field lines, and there are beams in both the outward (away from the 
neutron star) and inward directions, because the accelerating gaps are 
populated by pair production. They considered the pulse profile of 
radiation produced in the outer gap and showed that a single pole will 
produce a broad, irregularly-shaped, emission which is particularly 
dense near the edge. As a result, double $\gamma$-ray pulses will be 
observed when the line of sight from the Earth crosses these enhanced 
regions of the $\gamma$-ray beam, while the inner region of the beam 
provides a significant amount of emission between the pulses. With a
proper choice of the observer viewing angle, a wide range of peak phase
separations can be accommodated. Furthermore, Chiang and Romani 
(1994) refined the calculation of high-energy emission from the 
rotation-powered pulsars based on the CHR model. Their major refinements 
include (i) the approximate location of the emission at each point 
in phase along a given line of sight was inferred by using a pulse phase 
map, and (ii) because the spectral emissivities at different emission points 
will differ, so the outer gap is divided into small subzones in the 
plane containing the rotation and dipole axes. The photon densities and
beaming directions for different zone are also different, in which case 
the particle transport needs to be considered. Under their
refinements, they found that the spectral variation of the 
$\gamma$-radiation over the pulsar period is the result of the different
emission processes which play a role throughout the outer magnetosphere,
however, they were not able to obtain a self-consistent spectrum which
resembled the observed high-energy spectra, and they attributed this
shortcoming to the inability to model appropriately the extremely
complex emission processes and their interactions. Subsequently, Romani and his co-workers (Romani and Yadigaroglu, 1995; Yadigaroglu and Romani, 1995; Romani, 1996)  have improved their three dimensional models and successfully explain the high-energy emission features of pulsars including the phase-resolved spectra of the Vela pulsar. However, in their model they have assumed that there is only one single outer gap and only outgoing current. These two assumptions do not have real physical justification.

\section{CRZ MODEL}

Cheng et al.~(2000 hereafter CRZ) re-consider the three
dimensional magnetospheric outer gap model, following
the important ground-breaking work of Romani and co-workers. But instead of
assuming a {\it{single}} outer gap with {\it{only}} an outgoing current,
and 
no restriction on azimuthal directions, they use various physical 
processes (including pair
production which depends sensitively on the local electric field and the
local radius of curvature, surface field structure, reflection of
$e^{\pm}$ pairs because of mirroring and resonant scattering) to determine
the three-dimensional geometry of the outer gap. In their model, two outer
gaps and both outgoing and incoming currents are in principle allowed, but it
turns out that outgoing currents dominate the emitted radiation intensities. 
Furthermore, the three dimensional structure of outer gaps is completely determined by pair production 
conditions. Since the potential drop of the gap is
$\Delta V\approx 6.6\times 10^{12}f^2_0B_{12}P^{-2}\;{\mbox{Volts}}$,
where $P=2\pi /\Omega $ is the rotation period, $\Omega$ is the rotation angular velocity, $B_{12}$ is the surface magnetic field in units of 10$^{12}$ Gauss, $f_0=h(<r>)/R_L$, $h(<r>)$ is the average vertical separation of the
gap boundaries in the (${\bf{\Omega}}$, ${\bf{\mu}}$) plane and
$R_L=c/\Omega$ is the light cylinder radius, and $<r>$ is the average 
distance to the gap; its value depends on magnetic inclination angle 
$\alpha$ ($<r>\sim R_L/2$). The particle current passing through the gap is   
$\dot{N}_{gap}=3\times 10^{30}f_0\xi B_{12}P^{-2}\;{\mbox{s$^{-1}$}}$,
where $\xi=\Delta\Phi/2\pi$; $\Delta\Phi$ is the transverse
($\phi$-direction) extension of the gap. Each of the
charged particles inside the gap will radiate high-energy curvature
photons with a characteristic energy
$E_{\gamma}(f_0)=2\times 10^8 f^{3/2}_0B^{3/4}_{12}P^{-7/4}\;{\mbox{eV}}$.
About half of $\dot{N}_{gap}$ will move toward the star. Although they
 continue to radiate their energies on the way to the star, they
still carry 10.5$P^{1/3}$ ergs of energy on to the stellar surface.
The energy will be radiated back out in hard X-rays. However, 
resonant scattering with pairs near the star may reflect  hard
X-rays back to the stellar surface (Cheng et al.~1998; Wang et 
al. 1998), to be re-emitted as soft X-rays with a temperature 
$T_s\approx 3.8\times 10^6f^{1/4}_0\xi^{1/4}B^{1/2}_{12}P^{-5/12}\;{\mbox{K}}$.

The X-ray photon density is very low but each pair produced by an
X-ray-curvature photon collision in the outer gap will emit almost $10^5$ 
curvature $\gamma$-rays for further pair creation in that gap. Once the 
pair production threshold condition $kT_sE_{\gamma}\ge(m_ec^2)^2$
is satisfied, the gap is unlikely to grow much larger. This pair production
condition gives
$f_0=5.5P^{26/21}B^{-4/7}_{12}\xi^{1/7}$.
Here, $\xi$ is still an unknown quantity. However, $f_0$ 
is weakly dependent on $\xi$ which is likely of order of unity. In 
first approximation, they assume $f_0=5.5P^{26/21}B^{-4/7}_{12}$ (Zhang and
Cheng, 1997). To determine $\xi$, they consider local pair production
processes. The pair production per unit length inside the
gap is a decreasing function of $r$. According to CHR model, $E_{||}\propto r^{-1/2}$ for the thin outer gap (e.g.~the Crab 
pulsar), which gives $E_{\gamma}(r)\propto
r^{-1/8}$ after using the large $r$ limit $s(r)=(rR_L)^{1/2}$. Since
$E_{\gamma}$ is only weakly dependent on $r$, they assume
$\sigma_{\gamma\gamma}\approx const$.
The local pair production per unit length is 
$N_{e^{\pm}}(r)=(1-e^{-\tau_{\gamma\gamma}})N_{\gamma}(r)
\approx \tau_{\gamma\gamma}N_{\gamma}(r)$,
where $\tau_{\gamma\gamma}=n_X(r)\sigma_{\gamma\gamma}l(r)$ is the local
optical depth, $n_X=R^2T^4_s\sigma/r^2kT_sc$ is the X-ray number
density at $r$, $l(r)\approx (2s(r)f(r)R_L)^{1/2}$ is the local optical
path, $f(r)=h(r)/R_L$ is the local vertical extension of the gap (since
$B(r)h^2(r)$ is a constant, which gives $f(r)\propto r^{3/2}$ and $f_0\sim
f(R_L/2)$ ), and $N_{\gamma}=eE_{||}(r)/E_{\gamma}(r)$ is the number of
curvature photons emitted at $r$ per $e^+/e^-$ per unit length. Then
$N_{e^{\pm}}(r)\propto r^{-11/8}$.
Since most pairs are produced near the null surface where
$r=r_{in}$, so the pair production  take
place mainly in the range $r_{in}\leq r\leq r_{lim}$ where $r_{lim}$ 
is estimated as 
$r_{lim}N_{e^{\pm}}(r_{lim})/r_{in}N_{e^{\pm}}(r_{in})\sim
(r_{lim}/r_{in})^{-3/8}\sim 1/2$, which gives $r_{lim}\sim 6r_{in}$.		
 This limits pair production both along the field lines and in transverse 
 directions, and gives
$\Delta\Phi\sim 160^{\circ}$ by using the parameters of the Crab pulsar.

Within the pair production regions, outgoing and
incoming directions for particle flows are allowed. For $r>r_{lim}$
 only outgoing current is possible. Figure 1 shows our 3D
outer gap structure. Two pencil beams represent radio beams. The
light budge grey
shadow is the surface of last closed magnetic field lines, two dark
surfaces represent the upper boundaries of two outer gaps. So the
structure of the outer gap starts from the null surface and end at the
light cylinder. The lower boundary is last closed field surface and the
upper boudary is shown in Figure 1.
\begin{figure}[ht]
\begin{minipage}{85mm}
\includegraphics[width=75mm]{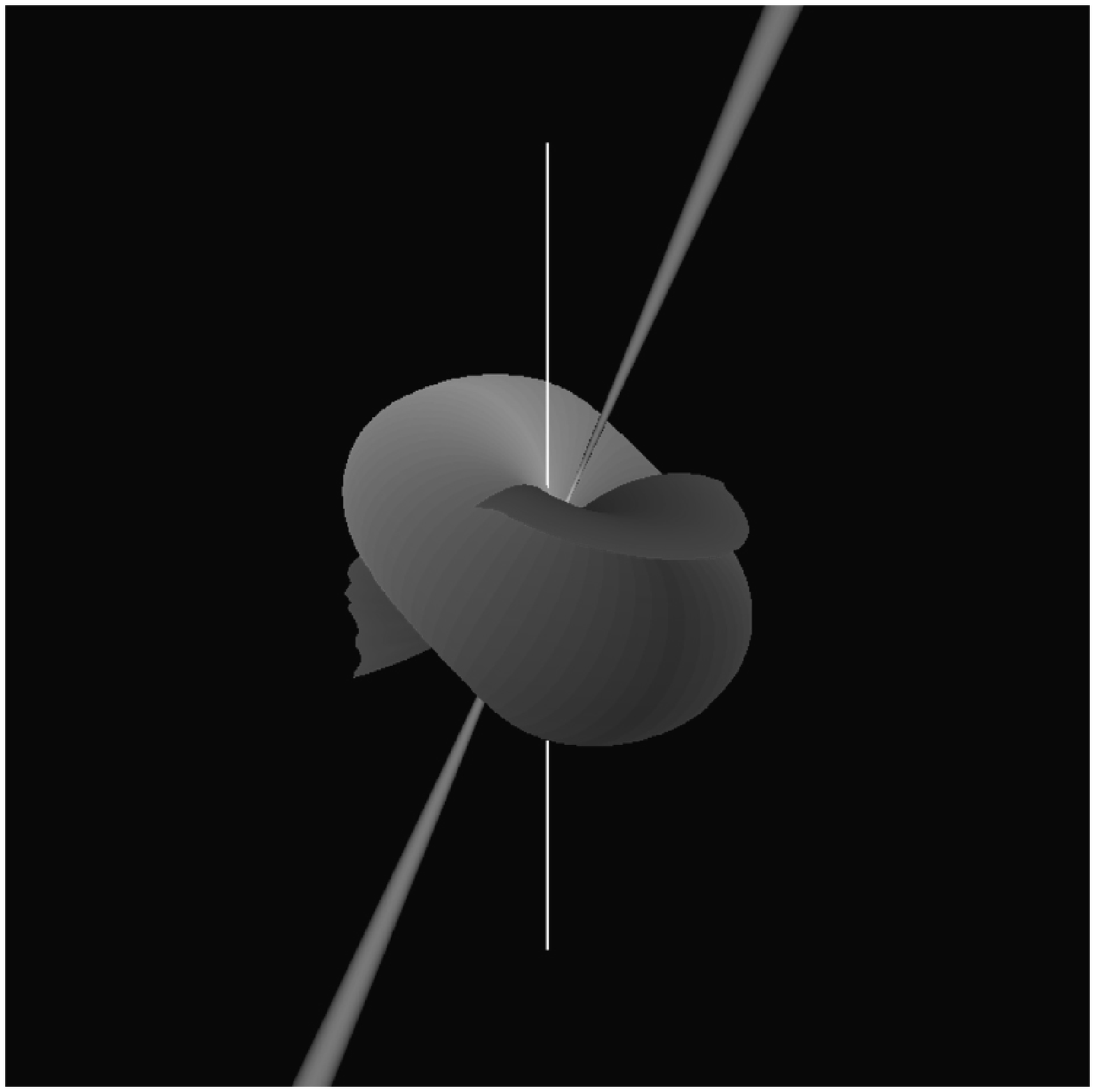}
\vspace{5mm}

{\sf Fig.~1.  3D structure of the outer-magnetospheric gaps.\label{fig:f1}}
\end{minipage} \hfil\hspace{\fill}
\begin{minipage}{85mm}
\hspace{-26mm}
\includegraphics[width=110mm]{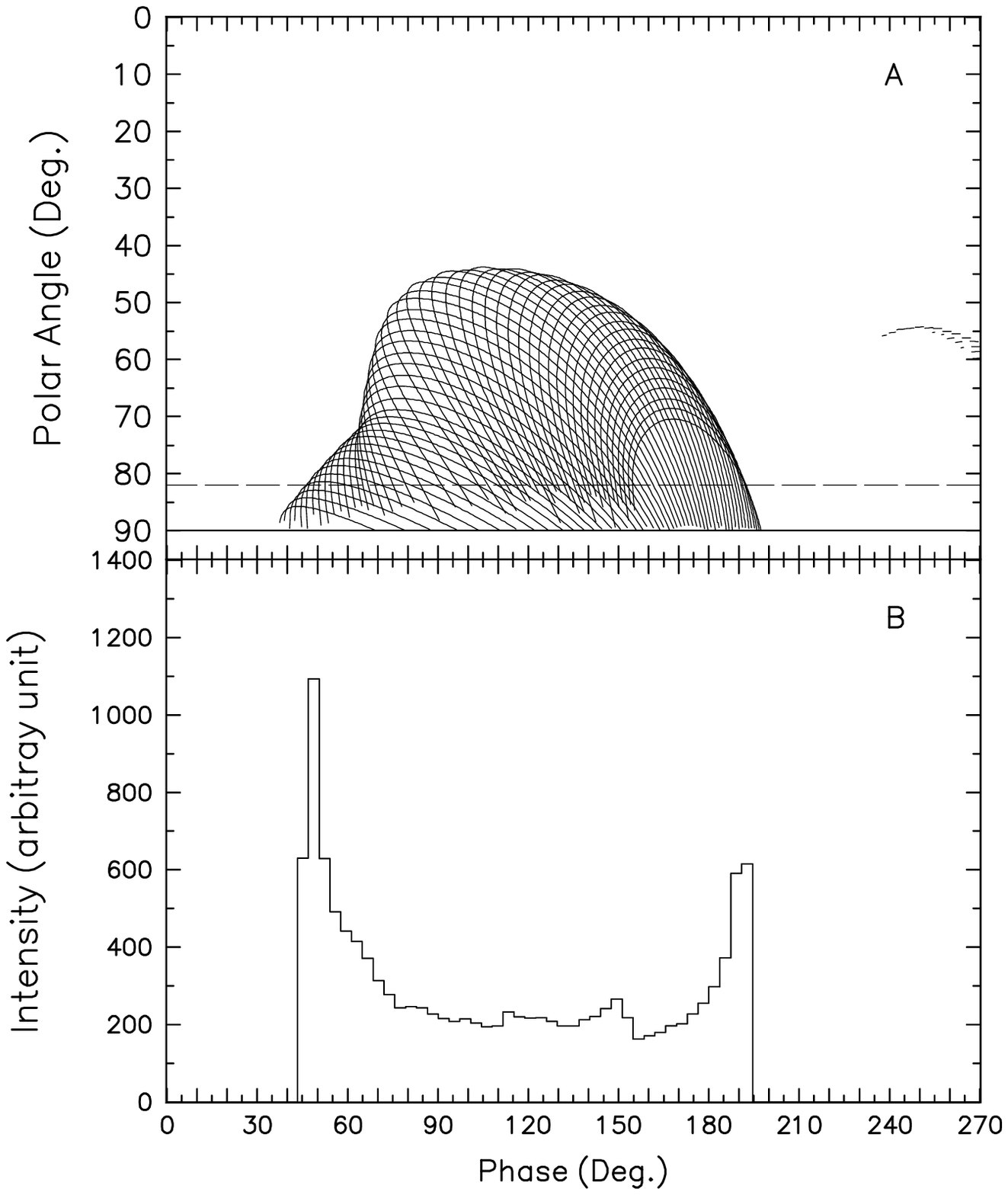}
\vspace{-39mm}

{\sf Fig.~2.  Emission projection onto the ($\zeta $, $\Phi $) plane and pulse profile. The emission consists of the
emission
outwards from both outer gaps and inwards only from the region ($r_{lim}-r_{in}$) of both outer gaps. The outer gaps are
limited along the azimuthal direction by pair production. Crab parameters, $\alpha=65^\circ $, $\zeta =82^\circ $ and $\Delta \Phi =160^\circ $
are used.\label{fig:f2}}
\end{minipage}
\end{figure}

%

\section{EMISSION MORPHOLOGIES AND LIGHT CURVES}

In this section, we discuss the morphological features of emission from the three dimensional outer gaps. It is important to note that in order to properly describe the three-dimensional pulsar magnetosphere three-dimensional rotating dipolar magnetic field must be used. 
Chiang and Romani (1994) and Romani and Yadigaroglu (1995) assumed the entire locus of points on the
last closed surface bounded by the null surface and the light cylinder as
giving emission in an outer gap. In our model, the emission-producing
outer gaps are limited along both radial and the $\phi$-directions. For
example, the extension of the outer gap on the $\phi$-direction is about
160$^{\circ}$ for Crab parameters with $\alpha=65^{\circ}$. We describe
the new photon emission morphologies below.

High-energy photons will be emitted nearly tangent to the magnetic field 
lines in the corotating frame 
because of the relativistic $1/\gamma$ beaming inherent in high 
energy processes unless $|\bf{E}\times\bf{B}|\sim B^2$. Then
following Romani and Yadigaroglu (1995), we assume
relativistic charged particles in the open zone radiate in their 
direction of propagation, i.e.~along the magnetic field lines in the
corotating frame. For each location within the open zone the direction of
emission expressed as ($\zeta$,$\Phi$) is calculated, where $\zeta$ is the
polar angle from the rotation axis and $\Phi$ is the phase of rotation of
the star. Effects of the time of flight and aberration
are taken into account. A photon with velocity ${\bf{u}}=(u_x,u_y,u_z)$
along a magnetic field line with a relativistic addition of velocity
along the azimuthal angle gives an aberrated emission direction
${\bf{u'}}=(u'_x,u'_y,u'_z)$. The time of flight gives a change of the phase of
the rotation of the star. Combining these two effects, and choosing
$\Phi=0$ for radiation in the (x,z) plane from the center of the star,
$\zeta$ and $\Phi$ are given by 
$\cos\zeta = u'_z$ and 
$\Phi =-\phi_{u'}-{\vec{r}}\cdot{\hat{u'}}$,
where $\phi_{u'}$ is the azimuthal angle of $\hat{u'}$ and $\vec{r}$ is 
the emitting location in units of $R_L$. We project photon 
emissions on the ($\zeta$, $\Phi$) plane and observe the emission 
patterns on the sky. In ($\zeta$, $\Phi$) plane, the null surface can be
determined easily because it consists of the points at which magnetic
field lines are perpendicular to the rotation axis. For a field line, the
null charge crossing is where the projected line crosses the
equatorial  line ($\zeta=90^{\circ}$). As an example, we show the
projections of photon emission both outwards and inwards from one outer
gap in ($\zeta$, $\Phi$) plane for the magnetic inclination angle
($\alpha=65^{\circ}$) in the upper panel of Figure 2, where the extension of the outer gaps on the
$\phi$-direction
is assumed to be $180^{\circ}$ and $(r_{lim}-r_{in})\sim 0.55R_L$(CRZ).
As mentioned above, the pulse profile depends on emission location and
viewing angle. Here we use Crab parameters and the viewing angle $\zeta=82^{\circ}$ ( the dashed line in panel A of Figure 2) to construct the light curve. The intensity of the light curve at each phase (panel B of
Figure 2) is
proportional to the number of interceptions between the dashed line and the solid lines in panel A of Figure 2.

%

\section{PHASE-RESOLVED SPECTRA OF THE CRAB PULSAR}

In this section, we describe how to calculate the phase-dependent spectra of pulsars. We will use the parameters of the Crab pulsar as example. Because the Crab pulsar outer gaps are thin, we use the electric
field of the CHR model:
\begin{equation}
E_{||}(r)={\Omega B(r)a^2(r)\over c s(r)}= {\Omega B(r)f^2(r)R^2_L\over c
s(r)},
\end{equation}
where $a(r)$ is the thickness of the outer gap at position $r$,  the radius of the curvature 
$s\sim (rR_L)^{1/2}$, and $f(r)\equiv a(r)/R_L$
 is
the local fractional size of the outer gap. Since the magnetic flux subtended in the
outer gap should be constant in the assumed steady state,
\begin{equation}
f(r)\sim f(R_L)\left({r\over R_L}\right)^{3/2},
\end{equation}
where $f(R_L)$ is estimated by the pair production condition described
in previous section(N.B.~There are two possible ways to determine $f(R_L)$, Eq. (6.7) of
CHR II or Eq. (22) of Zhang and Cheng (1997), but they come out very close to each other.). The local Lorentz factor of the
accelerated electrons/positrons in the outer gap is 
\begin{equation}
\gamma_e(r)=\left({3\over 2}{s^2\over e^2c}eE_{||}(r)c\right)^{1/4}.
\end{equation}
Because of the high soft photon density, the high-energy emission from
the Crab pulsar is described by
synchrotron self-Compton process. 

In an outer gap, the number of primary charged particles in a volume
element $\Delta V$ is roughly given by
$dN=n_{GJ}\Delta A\Delta l$,
where $n_{GJ}={\bf{\Omega}}\cdot{\bf{B}}/2\pi e c$ is the local
Goldreich-Julian number density, $B\Delta A$ is the magnetic flux through
the accelerator and $\Delta l$ is the path length along its magnetic field
lines. Using the thin gap approximation, the total number of charged
particles in the outer gap is 
\begin{equation}
N\sim {\Omega \Phi\over 4\pi c e}R_L,
\end{equation}
where $\Phi \sim f(R_L)B(R_L)R^2_L\Delta \phi$ and $\Delta \phi$ is the
angular range of the outer gap extending along the azimuthal 
direction, estimated in previous section. These primary $e^{\pm}$ pairs
will lose their energy by radiating curvature photons with a
characteristic energy
\begin{equation}
E_{cur}(r)={3\over 2}\hbar\gamma^3_e(r){c\over s(r)}.
\end{equation}
The power into curvature radiation for $dN$ $e^{\pm}$ pairs through in
A unit volume is
\begin{equation}
{dL_{cur}\over dV}\approx l_{cur} n_{GJ}(r),
\end{equation}
where $l_{cur}=eE_{||}(r)c$, is the local power into the curvature
radiation from a single electron/positron. The spectrum of primary
photons from a unit volume is 
\begin{equation}
{d^2\dot{N}\over dVdE_{\gamma}}\sim {l_{cur}n_{GJ}\over E_{cur}}{1\over
E_{\gamma}}
\end{equation}
where $E_{\gamma}\leq E_{cur}$.
These primary curvature photons collide with the soft photons produced by
synchrotron radiation of the secondary $e^{\pm}$ pairs, and produce the
secondary $e^{\pm}$ pairs by photon-photon pair production. Although
pair production inside an outer gap is limited to a small region
($r_{in}\leq r \leq r_{lim}$), pair production outside the outer gap can 
cover a much wider range because the synchrotron photons produced
by the secondary pairs are more abundant than the thermal photons from the
stellar surface. The former  cannot get into the outer gap because of the
field line  curvature (cf.~CHR I) but they can convert most curvature
photons from the outer gap into the secondary pairs. In a steady
state the distribution of secondary electrons/positrons in a unit volume
\begin{equation}
{d^2N\over dVdE_e}\approx {1\over \dot{E}_e}\int 
{d^2\dot{N}(E'_{\gamma}=2E'_e)\over dVdE_{\gamma}}dE'_e
\sim {1\over \dot{E}_e}{l_{cur}n_{GJ}\over E_{cur}}\ln\left({E_{cur}\over
E_e}\right),
\end{equation}
with $\dot{E}_e$  the electron energy loss into synchrotron radiation,
\begin{equation}
\dot{E}_e=-{2\over 3}{e^4B^2(r)\sin^2\beta(r)\over m^2c^3}
\left({E_e\over mc^2}\right)^2,
\end{equation}
$B(r)$ is the local magnetic field and $\beta (r)$  the local 
pitch angle,
\begin{equation}
\sin\beta(r)\sim \sin\beta(R_L)\left({r\over R_L}\right)^{1/2}.
\end{equation}
$\sin\beta(R_L)$ is the pitch angle at the light cylinder.  
Then the energy distribution of the secondary electrons/positrons in volume
$\Delta V (r)$,
\begin{equation}
\left({dN(r)\over dE_e}\right)\approx {d^2N\over dVdE_e}\Delta V(r)
\sim {1\over \dot{E}_e}{l_{cur}n_{GJ}\Delta V(r)\over
E_{cur}}\ln\left({E_{cur}\over E_e}\right).
\end{equation}
The corresponding photon spectrum of the synchrotron radiation is
\begin{equation}
F_{syn}(E_{\gamma},r)={3^{1/2}e^3B(r)\sin\beta \over mc^2 h}{1\over
E_{\gamma}}\int^{E_{max}}_{E_{min}}\left({dN(r)\over
dE_e}\right) F(x)dE_e
\end{equation}
where $x=E_{\gamma}/E_{syn}$.
\begin{equation}
E_{syn}(r)={3\over 2}\left({E_e\over mc^2}\right)^2{h e B(r)\sin\beta 
(r)\over mc}
\end{equation}
is the typical photon energy, and $F(x)=x\int^{\infty}_xK_{5/3}(y)dy$, where
$K_{5/3}(y)$ is the modified Bessel function of order 5/3. Similarly, the
spectrum of inverse Compton scattered photons in the volume $\Delta V(r)$ is
\begin{equation}
F_{ICS}(E_{\gamma}, r)=\int^{E_{max}}_{E_{min}}
\left({dN(r)\over dE_e}\right)\left({d^2N_{ICS}(r)\over
dE_{\gamma}dt}\right)
dE_e,
\end{equation}
where
\begin{equation}
{d^2N(r)_{ICS}\over dE_{\gamma}dt}=\int^{\epsilon_2}_{\epsilon_1}
n_{syn}(\epsilon,r)F(\epsilon, E_{\gamma}, E_e)d\epsilon,
\end{equation}
and 
\begin{equation}
F(\epsilon, E_{\gamma}, E_e)={3\sigma_Tc\over 4 (E_e/mc^2)^2}{1\over
\epsilon}\left[2q\ln q +(1+2q)(1-q)+{(\Gamma q)^2(1-q)\over 2(1+\Gamma
q)}\right],
\end{equation}
with $\Gamma=4\epsilon(E_e/mc^2)/mc^2$, $q=E_1/\Gamma (1-E_1)$ with
$E_1=E_{\gamma}/E_e$ and $1/4(E_e/mc^2) < q <1$. The number density of the
synchrotron photons with energy $\epsilon$ is  
\begin{equation}
n_{syn}(\epsilon, r)={F_{syn}(\epsilon)\over cr^2\Delta\Omega},
\end{equation}
where $F_{syn}$ is the calculated synchrotron radiation flux, and
$\Delta \Omega$ is the usual beam solid angle.

Since the outer gap of the Crab pulsar is very thin, it is sufficient to use one representative surface to calculate the high-energy radiation.
For a given viewing angle not only the light curve can be determined but also the exact emission regions in the outer
gap are known. Figure 3 shows the emission trajectories in the outer gap.
Once the radial distances of the emission regions are
determined, the spectrum of photon emission can be calculated for a given
radial distance $r$. Figure~4 shows the comparison between the observed
phase-resolved spectra of the Crab pulsar and the calculated spectra in the energy range of 10 MeV to 10 GeV (Zhang et al.~2000). The more detailed
phase-resolved spectra only in the energy range of EGRET can be found in CRZ. Figure 5 show the model energy
dependent light
curves of X-rays in four different energy channels (Zhang and Cheng, 2001b). Figure 6 shows the phase dependent
spectral indices of X-rays in four different energy ranges.  
\begin{figure}[ht]
\vspace{-20mm}
\hspace{20mm}
\includegraphics[width=85mm,angle=-90]{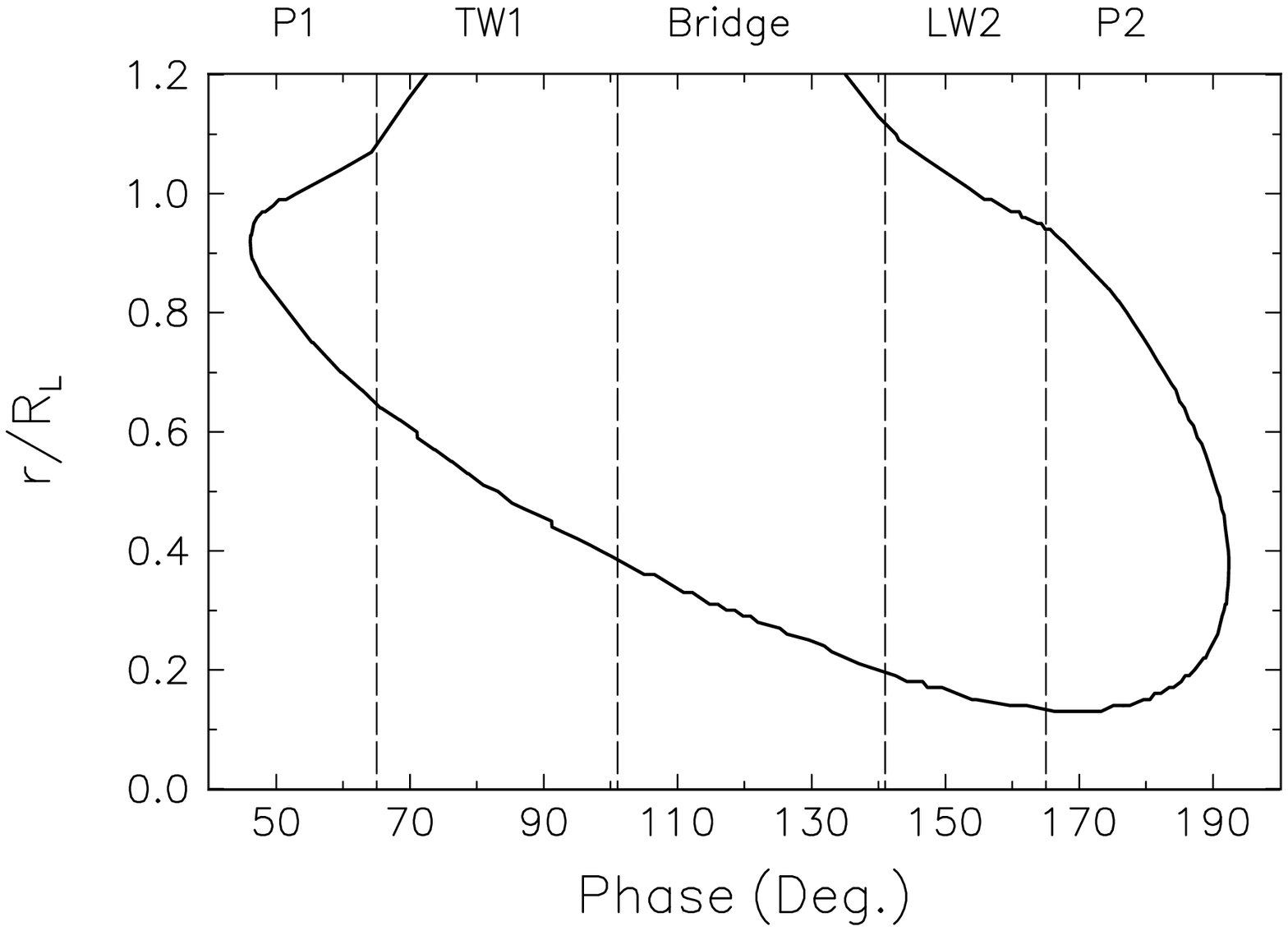}
\vspace{5mm}

{\sf Fig.~3.  The variation of radial distance with pulse phase for the Crab. The inclination angle is $65^\circ $. Five regions for different pulse phase are indicated.\label{fig:f3}}
\end{figure}
\begin{figure}[ht]
\hspace{-20mm}
\includegraphics[width=176mm]{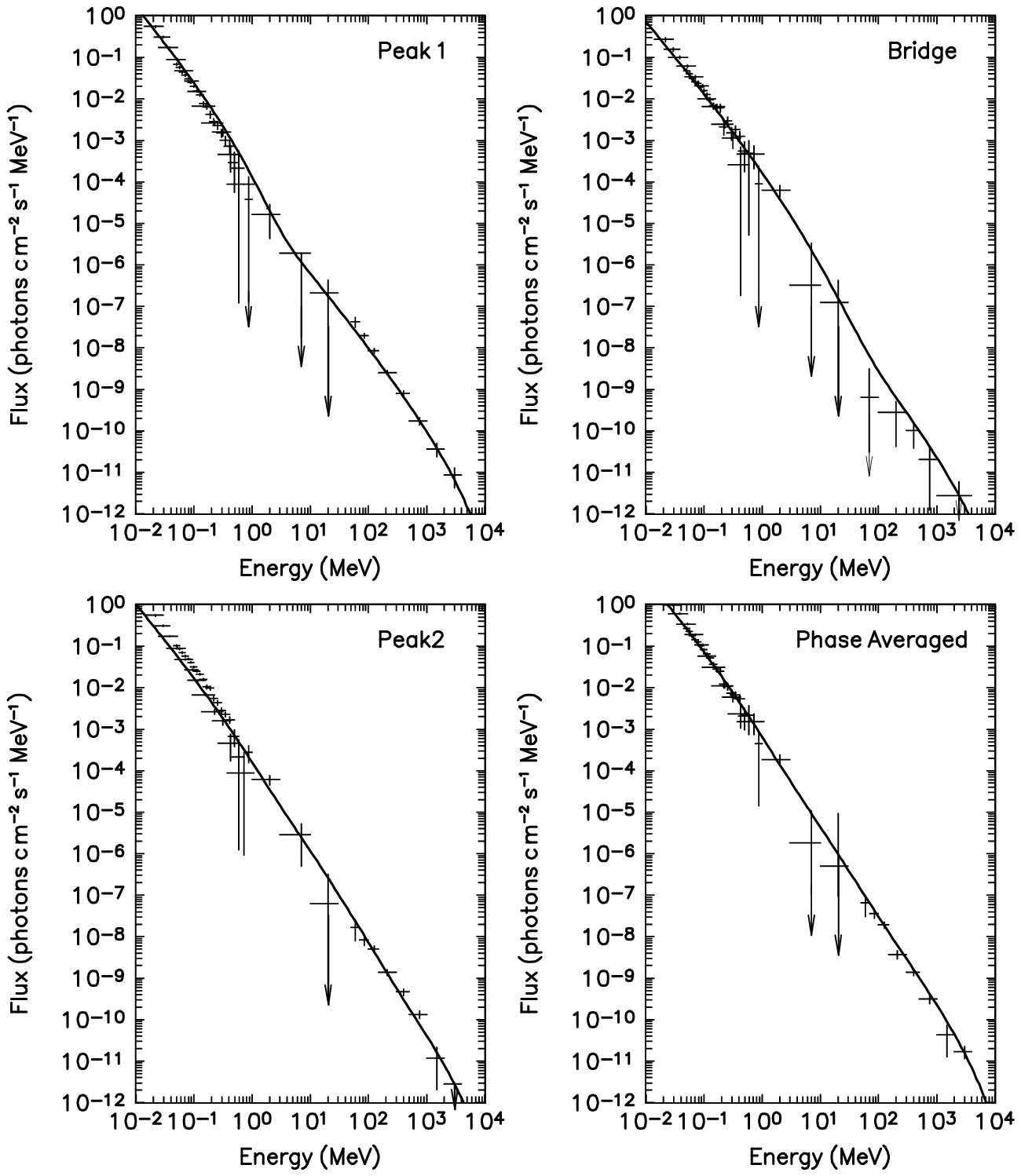}
\vspace{-55mm}

{\sf Fig.~4.  Phase-resolved $\gamma $-ray spectra from $10$MeV to $10$GeV for peak 1, bridge, peak 2 and phase-average
of the Crab pulsar. Observed data are taken from Ulmer et al.~(1995).\label{fig:f4}}
\end{figure}
\begin{figure}[ht]
\begin{minipage}{85mm}
\vspace{0mm}
\hspace{-30mm}
\includegraphics[width=110mm]{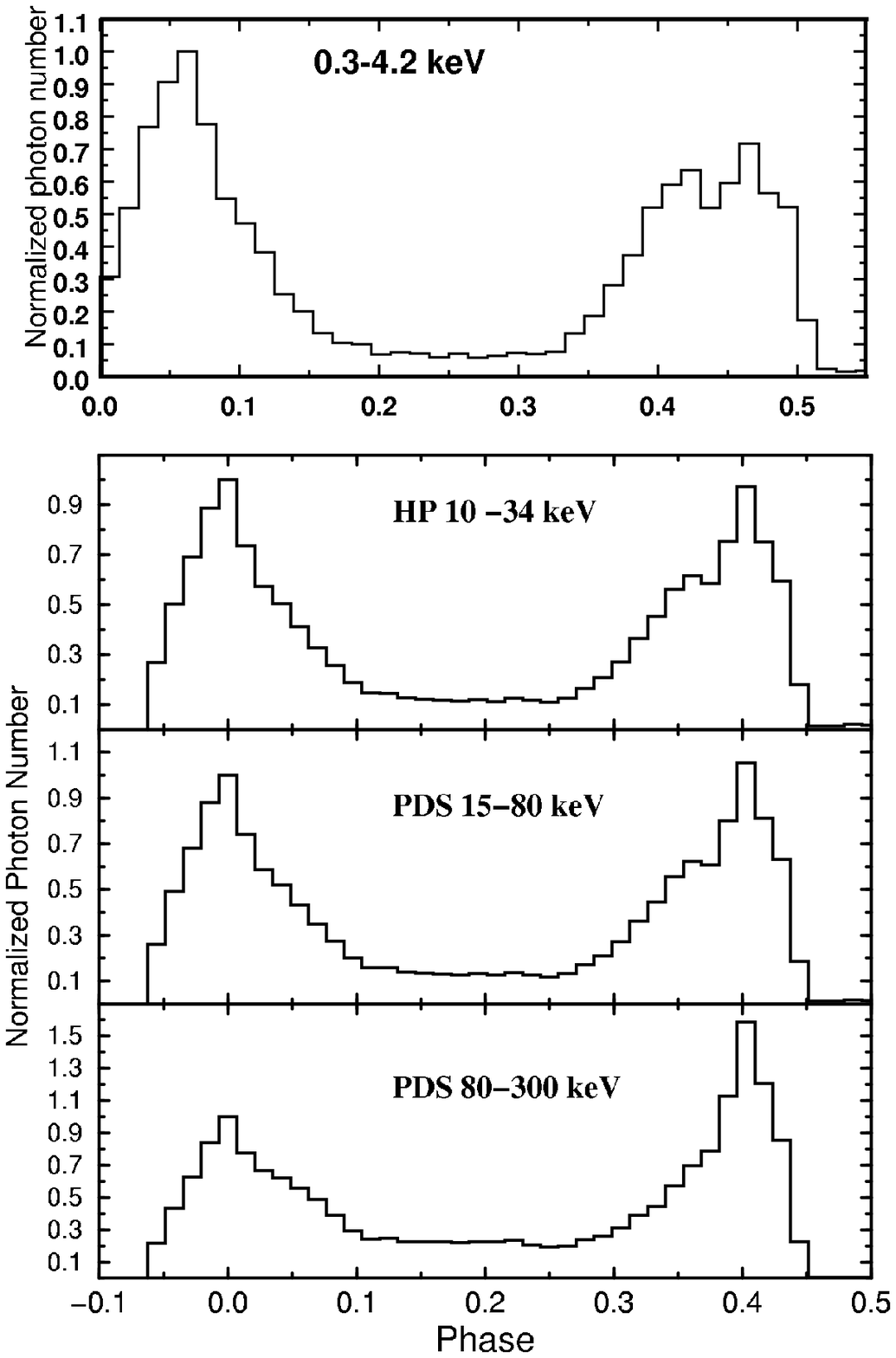}
\vspace{-30mm}

{\sf Fig.~5.  Expected X-ray pulse profiles of Crab pulsar for four different energy bands. The first model light curve
in the energy band from $0.3-4.2$ keV corresponds to the energy
range of the Chandra detector. The other three curves correspond to the energy ranges of the BeppoSAX
detectors. The magnetic inclination and viewing angles are assumed to be $65^\circ $ and $82^\circ $.\label{fig:f5}}
\end{minipage} \hfil\hspace{\fill}
\begin{minipage}{85mm}
\hspace{-30mm}
\includegraphics[width=110mm]{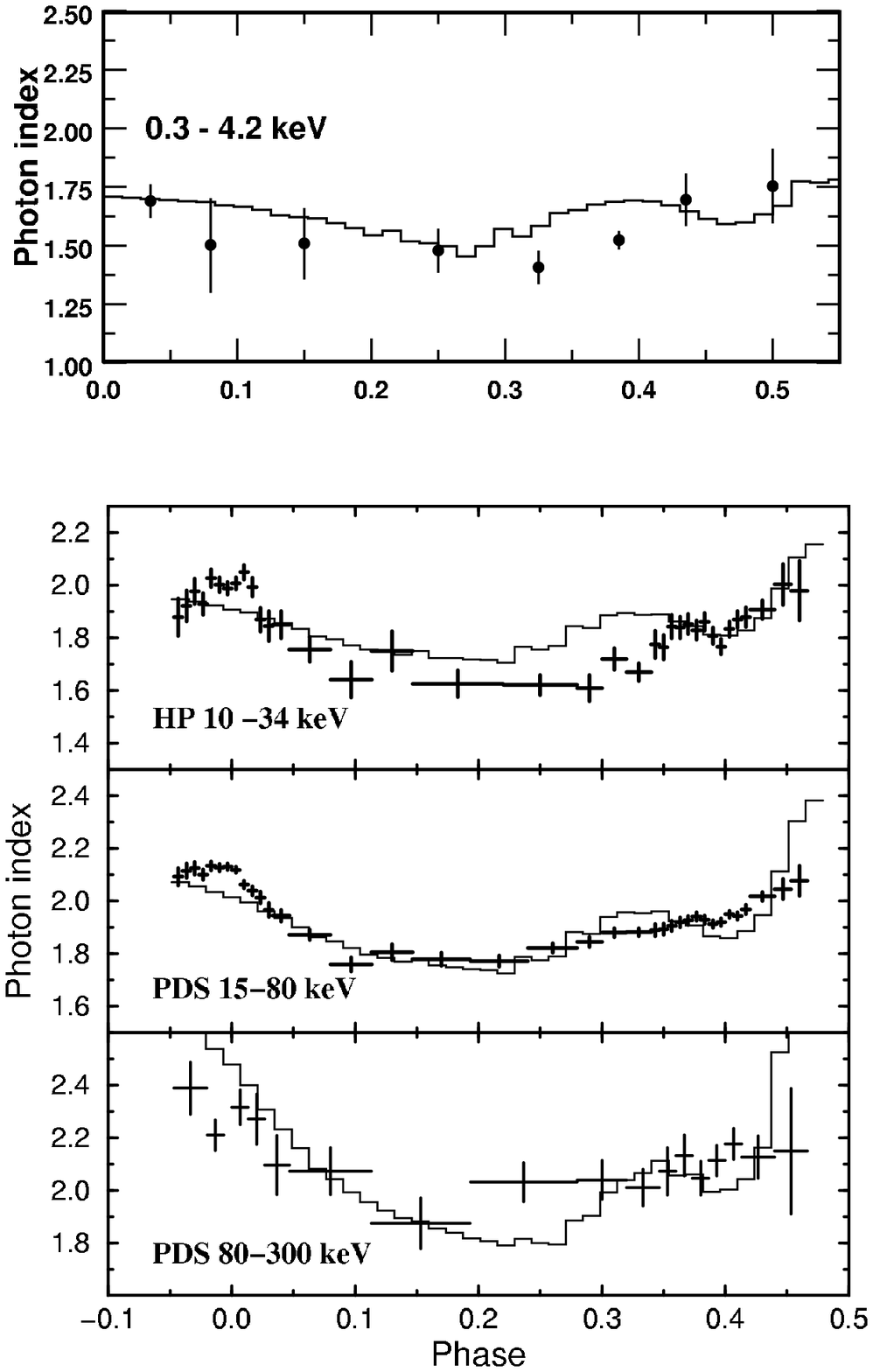}
\vspace{-30mm}

{\sf Fig.~6.  Comparison of expected spectral indices with the observed data. The magnetic inclination and viewing angles
are
assumed to be $65^\circ $ and $82^\circ $ for the Crab pulsar. The observed data are taken from Massaro et al.~(2000)
and Weisskopf (2002).\label{fig:f6}}
\end{minipage}
\end{figure}

\section{THE PHASE-RESOLVED SPECTRA OF THE GEMINGA PULSAR}

Fierro et al.~(1998) have shown the
observed light curve and phase-resolved spectra of high-energy
$\gamma$-rays of Geminga pulsar detected by EGRET. The observed pulse
profile by EGRET indicates that the phase separation is $0.49\pm 0.05$.
In order to obtain the observed phase-resolved
spectra, Fierro et al.~(1998) divided Geminga pulsar phase into 8 parts:
leading wing 1 (LW1), peak 1 (P1), trailing wing 1 (TW1), bridge, leading
wing 2 (LW2), peak 2 (P2), trailing wing 2 (TW2) and offpulse (OP). The
phase intervals widths of all these parts are 0.11, 0.09, 0.11, 0.15, 0.13, 0.13,
0.08 and 0.21 respectively. They have obtained the spectra for these different phase intervals and shown that the spectral indices change as a function of phase.

\begin{figure}[ht]
\vspace{-23mm}
\hspace{25mm}
\includegraphics[width=85mm,angle=-90]{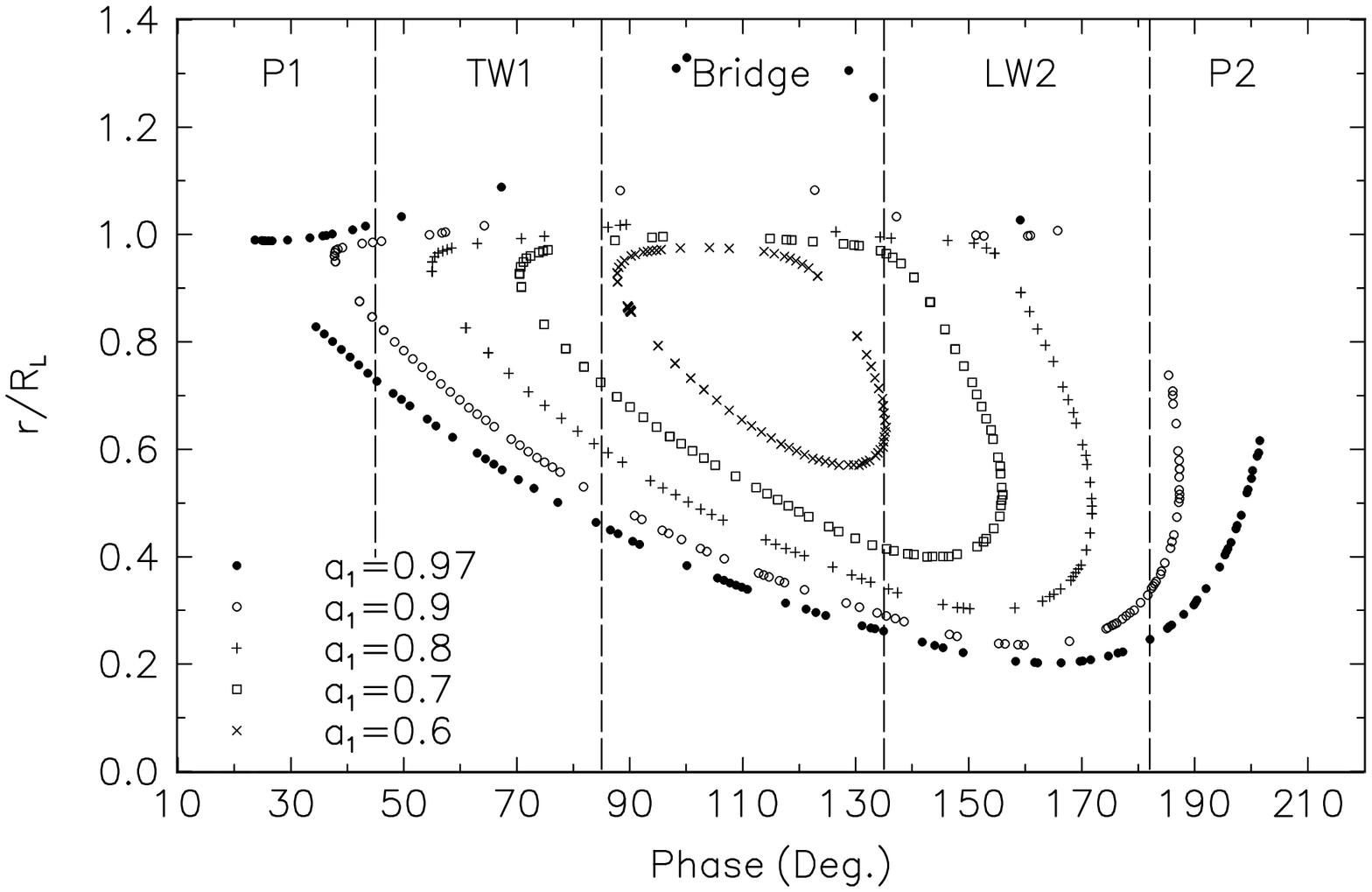}
\vspace{5mm}

{\sf Fig.~7a.  The variation of radial distance with the pulse phase 
for different outer gap surfaces of the Geminga pulsar. 
The inclination angle is $50^\circ $. Five regions for different pulse phases which are the same as observed those are
indicated.\label{fig:f7a}}
\end{figure}
\begin{figure}[ht]
\vspace{1mm}
\hspace{-10mm}
\includegraphics[width=160mm]{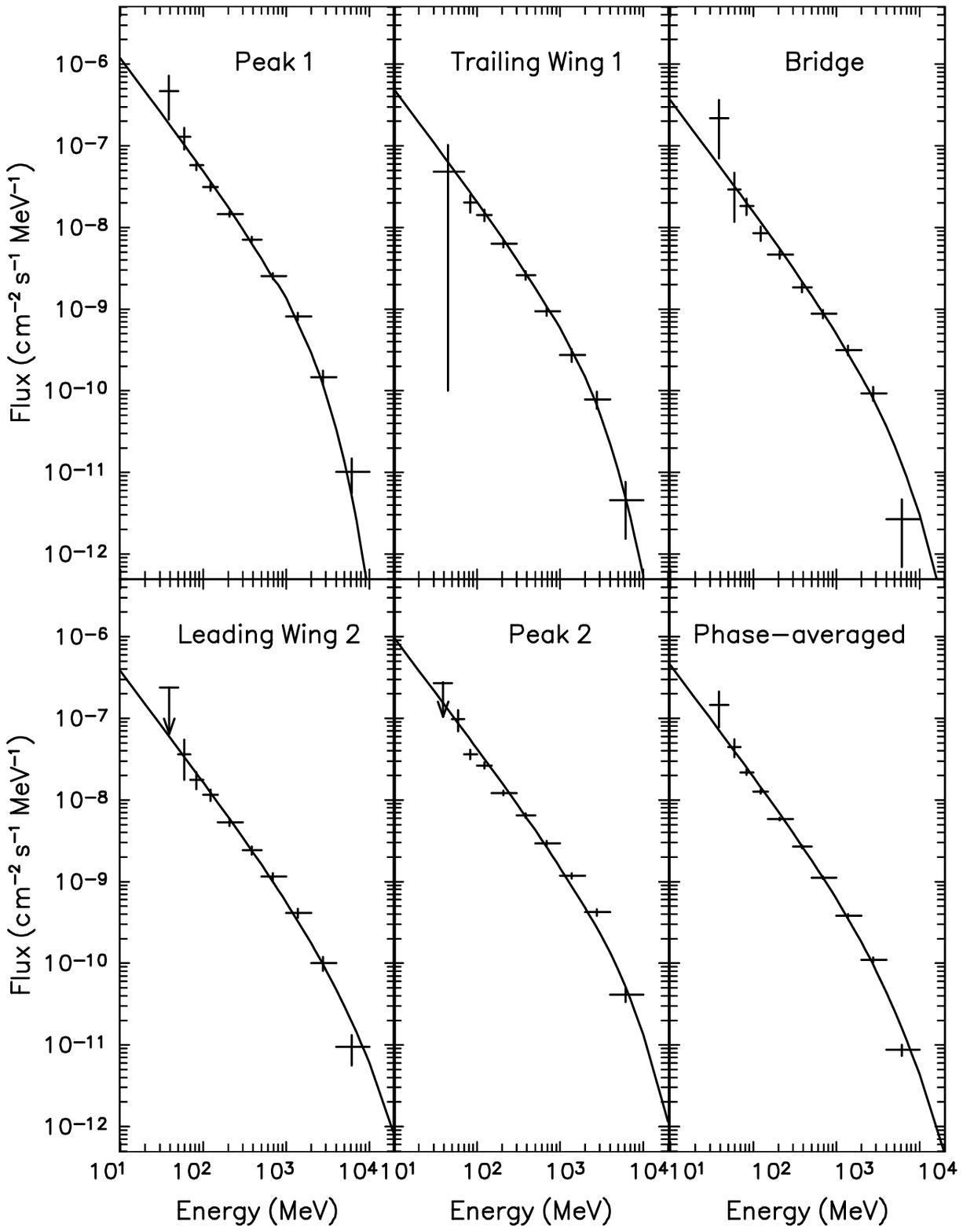}
\vspace{-45mm}

{\sf Fig.~7b.  Phase-resolved $\gamma $-ray spectra for different phases (peak one, trailing wing one, bridge, leading
wing two, peak 2 and phase-average) of the Geminga pulsar. Observed data are taken from Fierro et
al.~(1998).\label{fig:f7b}}
\end{figure}
Now we apply to this model to explain the pulse profile and phase-resolved
spectra of the Geminga pulsar. The key differences between the Geminga pulsar and the Crab pulsar are: (1) According to the model described in previous section, we can find that the fractional height of the outer gap for Geminga is
$f_0\sim 0.7$,which is a very thick gap. Certainly we cannot use a single surface to represent the emission regions. We
have used five layers to approximate the emission regions (cf.~Figure 7a). (2) The mean free path of the primary
photons
from the gap is longer than the light cylinder. So the observed $\gamma$-rays are curvature photons from primary charged
particles instead of synchrotron photons from the secondary pairs like in the case of the Crab pulsar. In order to
calculate the light curve and phase-resolved spectra, we also need to know $\alpha$ and $\zeta$. Since the Geminga
pulsar is a radio-quiet pulsar, these two parameters are difficult to know. Cheng and Zhang (1999) proposed a model for
X-ray emission from rotation-powered pulsars. They applied this model to the Geminga pulsar and found that the magnetic
inclination angle is $\sim 50^{\circ}$. Here, we use this value of the magnetic inclination angle. Furthermore the
$\gamma$-ray light curve of Geminga is nearly $\sim 180^{\circ}$ so the viewing angle must be very close to $\sim
90^{\circ}$, which is consistent to the fact that its radio beam cannot be within the line of sight. In Figure 7b,
we have compared the model phase-resolved spectra and the observed data.
The more detailed calculations on the X-ray and $\gamma$-ray light curves of the geminga pulsar can be found in Zhang and Cheng (2001a).

\section{APPLICATIONS TO OTHER CRAB-LIKE PULSARS: PSR B0540-69 AND PSR B1509-58}

We have also applied this three dimensional outer gap model to other Crab-like pulsars:PSR B1509-58 and PSR B0540-69, and model the light curves and the spectra
of optical, X-rays and $\gamma$-rays from these two pulsars (Zhang and Cheng, 2001b). Although the mean free path of
the primary photons from the outer gaps of these two pulsars is much shorter than the light cylinder and hence the radiation from
secondary pairs still dominate in the observed spectrum, the thickness of the outer gap in these two pulsars are 0.25
and 0.3 respectively. Again the emission regions cannot be approximated by a single surface. In Figure 8, we
have compared the model spectra and the observed spectra of PSR B0540-69 from optical to gamma-rays. The more detailed comparison between model results and the observed data of PSR B1509-58 and PSR B0540-69 can be found in Zhang and Cheng (2001b)
\begin{figure}[ht]
\vspace{-24mm}
\hspace{20mm}
\includegraphics[width=85mm,angle=-90]{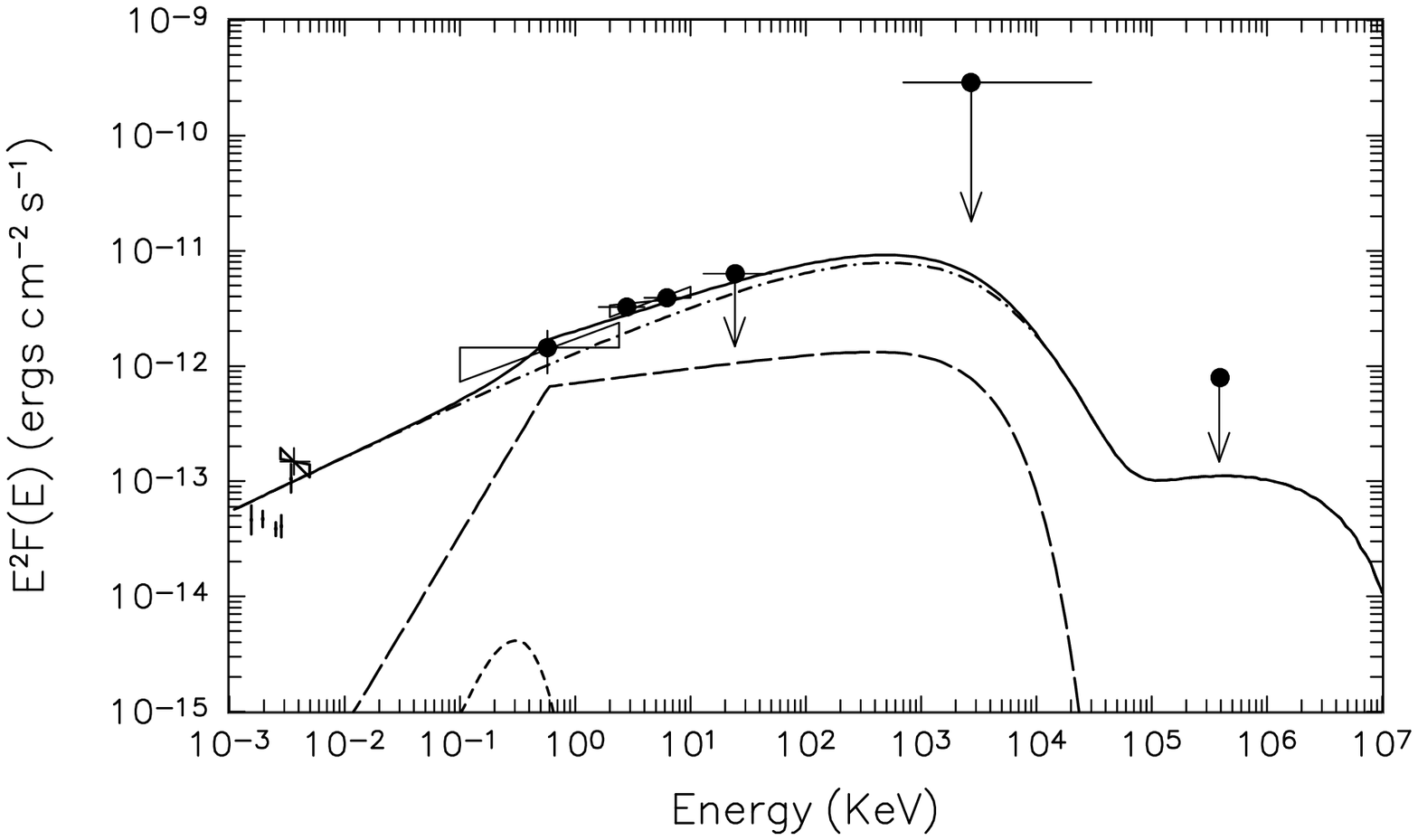}
\vspace{5mm}

{\sf Fig.~8.  The comparison of predicted phase-averaged spectrum with the observed data for
PSR B0540-69. Observed data at optical waveband are taken from Middleditch et al.~(1987), Hill et al.~(1997). 
The data at ROSAT energy range, BeppoSAX energy range, COMPTEL and EGRET are taken from
Finley et al.~(1993), Mineo et al.~(1999), Hermsen et al.~(1994) and Thompson et
al.~(1994) respectively. The solid curve represents phase-averaged spectrum. The dot-dashed, long-dashed and short-dashed curves represent
photon spectra in different phases.\label{fig:f8}}
\end{figure}

\section{SUMMARY AND DISCUSSION}

We use a 3-D model magnetosphere to model the observed light
curve and the phase-resolved spectra of the $\gamma$-ray pulsars. In our 
model, the local photon-photon pair production in the outer gaps limits 
the extension of the outer gaps along the azimuthal direction. We find 
that the two topological disconnected outer gaps, with some extension 
along the azimuthal ($\phi$) direction, exist in the pulsar 
magnetosphere. Double-peaked pulse profiles with varying phase 
separation, depending on viewing angle, and strong bridge emission 
occur naturally, as in the single pole outer gap model. In case of the Crab pulsar, pair production is not limited to inside the outer gap because the 
intense X-rays produced by secondary pairs in the outer-magnetosphere of
the Crab pulsar can convert most curvature photons into pairs outside
the gap. The observed spectrum of the Crab pulsar 
results from a synchrotron-self-Compton mechanism. We (Cheng et al.~2000) have tried to apply the 3D outer gap model
to explain the Crab pulsar's phase-resolved spectra and find some discrepancies between model results and the observed data. One possibility is that the data presented by Fierro et al.~(1998) contains the DC emission from the Crab/Nebula.

Although our three dimensional model can explain the observed phase-resolved spectra and the energy dependent light curves of various gamma-ray pulsars including the Crab pulsar and the Geminga pulsar reasonably well, the present model is subjected to number of limitations. First, the photons in the trailing wing 1, off-pulse and leading wing 1 in the light curve of the Crab pulsar (Fierro et al.~1998) cannot be explained by this model. Secondly, although the model Chandra light curve of the Crab pulsar and the observed one (cf.~Figure 4 of Weisskopf, 2002) are very similar,
we can only fit the phase-dependent spectral indices in half of the period (cf.~Figure 6). Again, we can explain the spectra of PSR B0540-69 and PSR B1509-58 reasonably well but the model X-ray light curves  are narrower than the observed ones.

In order to explain these discrepancies between the model results and the observed data, some improvements can be made in our model. (1) In our calculations, photons are assumed to be emitted tangent to the local field lines.
 Therefore, the emission light curves (cf.~Figure 3) have sharp edges in
both ends. In fact, some of secondary pairs have quite large pitch angles whose emission will not be tangent to the local field lines. Also for those pairs created near the null charged surface even they begins with small pitch angles, but when they stream towards the star and part of them will reflect back outwards due to the magnetic mirroring effect (Ho, 1988). Then they could end up with very large pitch angles. (2) The realistic magnetic field configuration in the pulsar magnetosphere could be different from a simply rotating dipolar magnetic field structure. (3) The particle energy density and the magnetic energy density are comparable near the light cylinder. Charged particles are not necessary straightly moving along the field lines. These pairs will have a much larger pitch angles and their radiation should not be restricted between pulses. (4) The charged current inside the outer-magnetospheric gap is assumed to be Goldreich-Julian current (1969), which must be the maximum value. In order to determine the real current flow inside the gap, it is necessary to solve a more consistent electro-dynamic model 
(Hirotani, 2002).

Finally, we would like to make a few remarks. The better test of the three dimensional outer gap model is to compare the model results with the phase-resolved spectrum from X-ray band to $\gamma$-ray band together, instead of comparing them separately. For the Crab pulsar, Kuiper et al.~(2001) have published a full observational (phase-resolved) picture from soft X-rays up to high-energy gamma-rays. For the Geminga Pulsar, Jackson et al.~(2002) have presented the combined data of ASCA, CGRO and RXTE. For other Crab-like pulsars, e.g.~PSR B1509-58 ( Kuiper et al.~1999; Cusumano et al.~2001) and PSR B0540-69 (de Plaa et al.~2003), more X-ray and $\gamma$-ray have been reported. We must carefully analyze these new data to see if they really support a simple three dimensional outer gap model. In fact, de Plaa et al.~(2003) have point out that there are noticeable discrepancy between latest data and model predictions (Cheng et al.~2000). They suggest that these discrepancies may result from the uncertainties in the pulsar geometry.

\section*{ACKNOWLEDGEMENTS}
We thank S.F. Ko, M. Ruderman and L. Zhang for useful discussion and suggestions. This work is partially supported by an RGC grant of the Hong Kong Government.

\end{document}